\documentclass[12pt]{article}
\usepackage{graphicx}
\usepackage{wrapfig}

\usepackage{caption}

\usepackage{lineno}

\usepackage[font=footnotesize]{caption}

\def\be{\begin{equation}}
\def\ee{\end{equation}}
\def\bea{\begin{eqnarray}}
\def\eea{\end{eqnarray}}

\def\mw{\ensuremath{m_{W}}}

\newcommand{\tch}{\textit{t}-channel}

\def\F0{\ensuremath{F_{\mathrm{0}}}}

\def\vl{\ensuremath{V_{\mathrm{L}}}}
\def\vr{\ensuremath{V_{\mathrm{R}}}}
\def\gr{\ensuremath{g_{\mathrm{R}}}}
\def\gl{\ensuremath{g_{\mathrm{L}}}}

\def\vlr{\ensuremath{V_{\mathrm{L,R}}}}
\def\glr{\ensuremath{g_{\mathrm{L,R}}}}

\newcommand{\Lagr}{\mathcal{L}}
\newcommand{\ProjR}{\ensuremath{P_{\mathrm{R}}}}
\newcommand{\ProjL}{\ensuremath{P_{\mathrm{L}}}}

\newcommand\pubnumber{SNSN-323-63}
\newcommand\pubdate{\today}

\def\institute{Instituto de f\'isica corpuscular (IFIC)\\
Centro mixto CSIC y Universitat de Val\`encia\\
Parque Cient\'ifico, C/Catedr\'atico Jos\'e Beltr\'an 2, E-46980 Paterna, SPAIN}
\def\support{\footnote{Work supported by the Spanish grants FPA2015-65652-C4-1-R, SEV-2014-0398 (MINECO), the Prometeo II/2014/016 (Generalitat Valenciana) and the MINECO PhD grant BES-2013-063738}}

\def\Title#1{\begin{center} {\Large #1 } \end{center}}
\def\Author#1{\begin{center}{ \sc #1} \end{center}}
\def\Address#1{\begin{center}{ \it #1} \end{center}}

\newcommand\pubblock{\rightline{\begin{tabular}{l} \pubnumber\\
         \pubdate  \end{tabular}}}
\newenvironment{Abstract}{\begin{quotation}  }{\end{quotation}}
\newenvironment{Presented}{\begin{quotation} \begin{center} 
             PRESENTED AT\end{center}\bigskip 
      \begin{center}\begin{large}}{\end{large}\end{center} \end{quotation}}

\def\beq{\begin{equation}}
\def\eeq#1{\label{#1}\end{equation}}
\def\eeqn{\end{equation}}

\def\beqa{\begin{eqnarray}}
\def\eeqa#1{\label{#1}\end{eqnarray}}
\def\eeqan{\end{eqnarray}}

\let\bar=\overbar

\def\Dslash{\not{\hbox{\kern-4pt $D$}}}
\def\dslash{\not{\hbox{\kern-2pt $\del$}}}

\def\ee{e^+e^-}

\def\mw{m_W}

\def\msb{{\bar{\ssstyle M \kern -1pt S}}}

\begin{document}
\begin{titlepage}
\pubblock

\vfill
\Title{Top quark property measurements in single top}
\vfill
\Author{ javier jim\'enez pe\~na\support \\ on behalf of the atlas and cms collaborations}
\Address{\institute}
\vfill
\begin{Abstract}
A review of the recent results on measurements of top quark properties in single top quark events using data samples of proton proton collisions produced by the LHC at $\sqrt{s}$ = 7 and 8 TeV and collected by ATLAS and CMS detectors is presented, including the top quark polarization, the W boson helicity, searches for anomalous Wtb couplings and the top quark mass.
The measurements are in good agreement with predictions and no deviations from Standard Model expectations have been observed.
\end{Abstract}
\vfill
\begin{Presented}
$9^{th}$ International Workshop on Top Quark Physics\\
Olomouc, Czech Republic,  September 19--23, 2016
\end{Presented}
\vfill
\end{titlepage}
\def\thefootnote{\fnsymbol{footnote}}
\setcounter{footnote}{0}

\section{Introduction}

At hadron colliders, top quarks are mainly produced in t\=t pairs through strong interaction.
However, a smaller fraction of top quarks are singly produced through weak interaction, involving the Wtb vertex. 
At LO, there are three sub-processes producing single top quarks: via the exchange of a virtual W boson in the t-channel or s-channel, or via the associated production of an on-shell W boson, in the Wt channel.

Single top quark topology opens a window to access the Wtb couplings, allowing searches of anomalous couplings through the production and decay mechanisms.

A review of the latest results on top quark properties for ATLAS \cite{ATLAS} and CMS \cite{CMS} is presented in this paper including: the top quark polarization, the W boson helicity, searches for anomalous Wtb couplings and the top quark mass.

All analysis presented in this paper use the 7 and/or the 8 TeV datasets from LHC \cite{LHC} data. ATLAS: 4.59 $\mathrm{fb}^{-1}$  at 7 TeV and 20.3 $\mathrm{fb}^{-1}$ at 8 TeV. CMS: 5.0 $\mathrm{fb}^{-1}$  at 7 TeV and 19.7 $\mathrm{fb}^{-1}$  at 8 TeV. 
All analysis use the leptonic decay of the W boson.

\section{Polarization and anomalous Wtb couplings}

\begin{wrapfigure}{r}{0.33\textwidth}
 \vspace{-30pt}
   \begin{center}
     \includegraphics[width=0.33\textwidth]{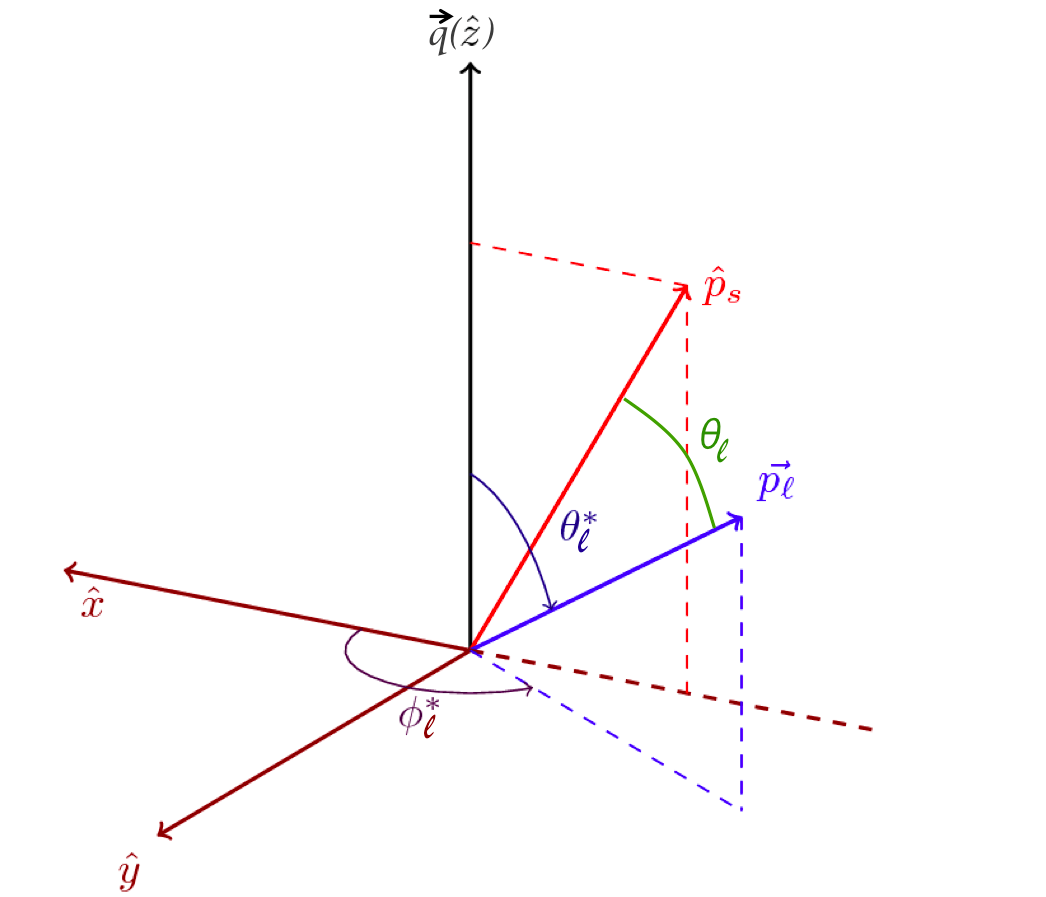}
    \end{center}
     \vspace{-20pt}
  \caption{Coordinate system and angles used to define the W boson spin observables and their related angular asymmetries in the decay of polarized top-quarks.}
 \vspace{-10pt}
\end{wrapfigure}

In the Standard Model (SM), the top quark is highly polarized in the direction of the
spectator quark in the \tch\ production.
Its spin is also correlated with the angular properties of the decay products. 
The angle $\theta_{l}$ between the charged lepton ($\vec{p_{l}}$) and non \textit{b}-tagged jet (i.e. spectator quark, $\hat{p_{s}}$) in the top quark rest frame defines the top quark polarization (Figure 1).

Within the SM, 
the top quark decays through the electroweak interaction, almost exclusively into a W boson and a b quark. 
The W boson also possesses a polarization that can be assessed via angular distributions of its decay products. 
For instance, the W boson helicity is defined by the angle $\theta^{*}_{l}$ between the W boson momentum in the top quark rest frame ($\vec{q}$) and the momentum of the lepton in the W boson rest frame ($\vec{p_{l}}$) and the azimutal angle  $\phi^{*}_{l}$.

The electroweak production and subsequent decay of single top quarks are determined by the properties of the Wtb vertex.
Deviations from the SM in the Wtb vertex can be expressed in terms of the
anomalous couplings, $\vlr$ and $\glr$, presented in this
effective Lagrangian:~\cite{Aguilar2008,Aguilar2010}:
\begin{equation}
\Lagr_{\mathrm{\it{Wtb}}} = - \frac{g}{\sqrt{2}}{\overline{b}}\gamma^\mu \left (\vl \ProjL + \vr \ProjR \right )tW^-_\mu - 
  \frac{g}{\sqrt{2}}{\overline{b}}\frac{i\sigma^{\mu\nu}q_{\nu}}{\mw}
  \left (\gl \ProjL + \gr \ProjR \right)tW^-_\mu + \mathrm{h.c.}
\end{equation}
Within the SM, $V_{L} = V_{tb} \approx 1$ and all the other couplings vanish at tree level, while they are non-zero at higher orders. 
The presence of non-SM couplings would modify the expected values for the top quark and W boson polarizations, and this feature has been exploited both by ATLAS and CMS in anomalous couplings searches.

\begin{wrapfigure}{r}{0.39\textwidth}
 \vspace{-20pt}
   \begin{center}
     \includegraphics[width=0.39\textwidth]{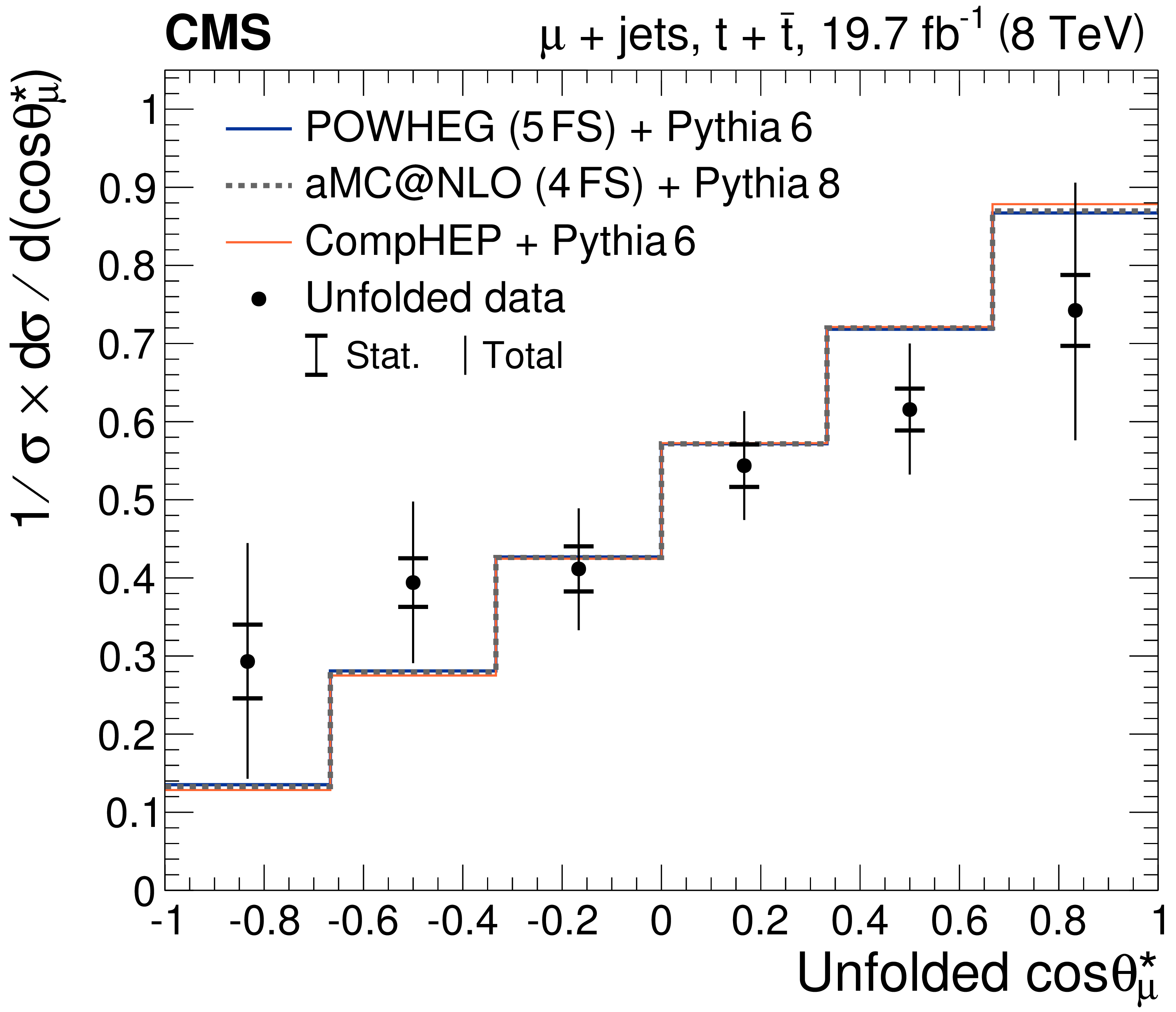}
    \end{center}
     \vspace{-16pt}
  \caption{The normalized differential cross section as a function of unfolded $cos(\theta_{\mu})$ compared to the theoretical predictions. \cite{CMSpol}}
 \vspace{-10pt}
\end{wrapfigure}

CMS has measured the top quark spin forward-backward asymmetry in the t-channel single top production using the 8 TeV data, in events with the W boson decaying to a muon and a neutrino \cite{CMSpol}.
Two consecutive BDT have been used in order to enhance the signal to background ratio. 
Background subtracted data is unfolded to parton level to correct for efficiency and detector resolution effects. 
The asymmetry is measured at parton level from the $cos(\theta_{\mu})$ distribution (Figure 2), obtaining $A^{\mu}_{\mathrm{FB}} = 0.26 \pm 0.03\,(\mathrm{stat.})  \pm 0.10\,(\mathrm{syst.})$, compatible with a p-value of 4.6\% with the SM expectation (0.44).

CMS has also studied, using the 8 TeV dataset, the W boson helicity fractions \cite{CMShel} .
The data is fitted to distributions of $cos(\theta_{l}^{*})$ built with signal plus background MC-reweighted samples.
Electron and muon channels are fitted independently and the results are later combined, obtaining $F_{\mathrm{L}} = 0.298 \pm 0.028\,(\mathrm{stat.})\pm 0.032\,(\mathrm{syst.})$,  $F_{\mathrm{R}} = 0.720 \pm 0.039\,(\mathrm{stat.})\pm 0.037\,(\mathrm{syst.})$ and $F_{\mathrm{0}} = -0.018 \pm 0.019\,(\mathrm{stat.})\pm 0.011\,(\mathrm{syst.})$. These results are later used to set limits on the real part of $g_{\mathrm{L}}$ and $g_{\mathrm{R}}$ (Figure 3).

\vspace{10pt}
\begin{centering}
\begin{minipage}[c]{0.54\textwidth}
\centering
\hspace{-10pt}
\includegraphics[width=0.6\textwidth]{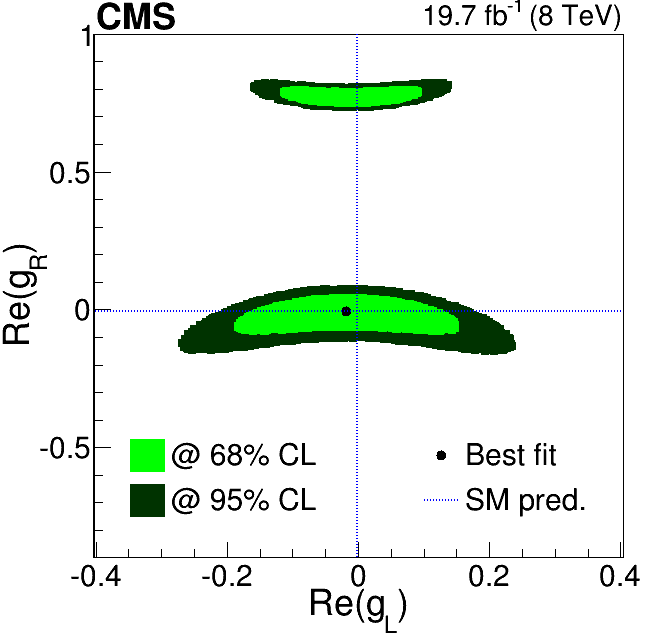}
\vspace{-5pt}
\captionof{figure}{Exclusion limits on the real part of $g_{L}$ and $g_{R}$ anomalous couplings, with $V_{L} = 1$ and $V_{R} = 0$. \cite{CMShel}}
\end{minipage}
\hspace{10pt}
\begin{minipage}[c]{0.42\textwidth}
\centering
\hspace{-20pt}
\includegraphics[width=1.031\textwidth]{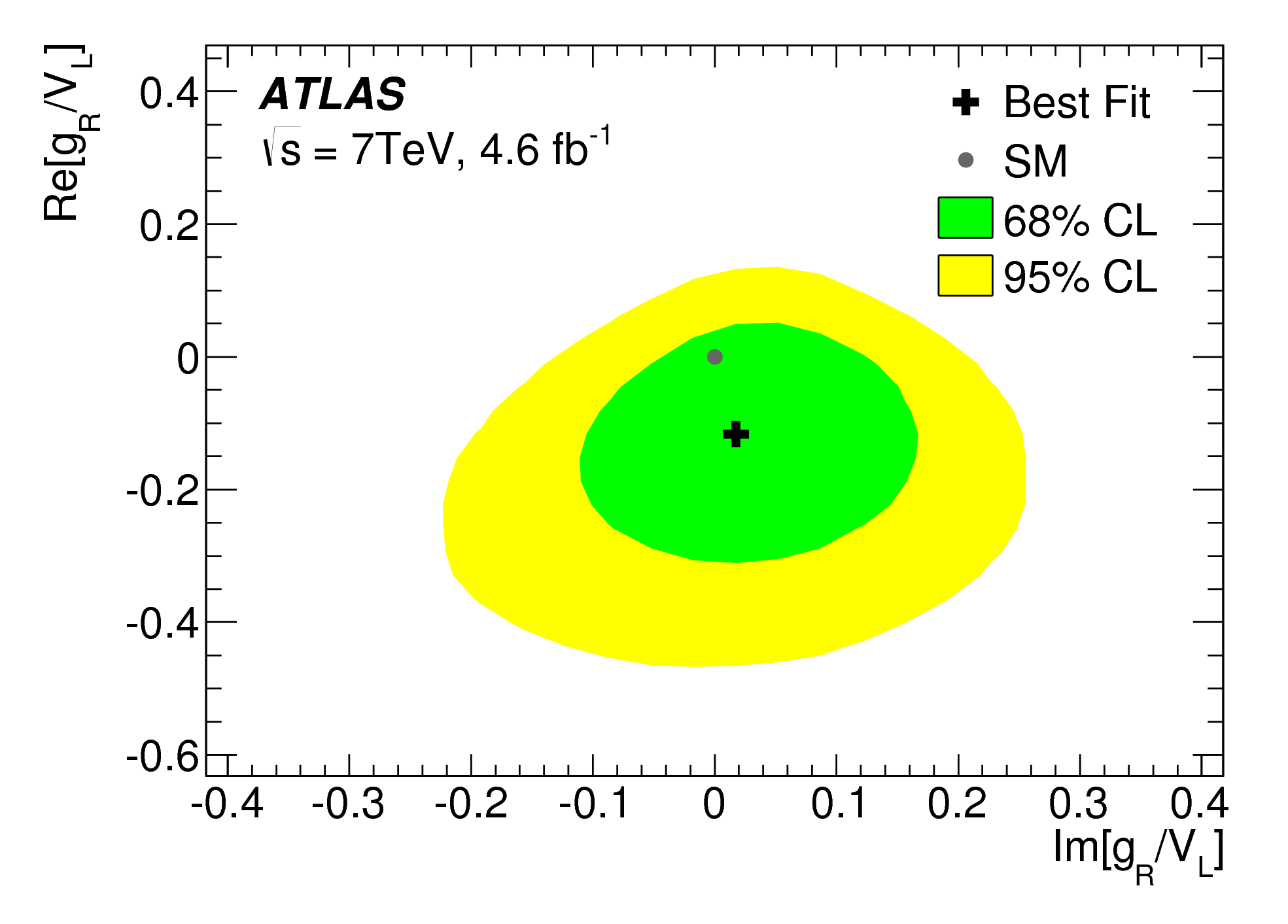}
\vspace{-5pt}
\captionof{figure}{Exclusion limits on $Re[g_{R}/V_{L}]$ vs $Im[g_{R}/V_{L}]$, with $V_{R} = g_{L} = 0$. \cite{ATLASangles} }
\end{minipage}
\end{centering}

An analysis in ATLAS, using 7 TeV data, performs a normalized double differential angular measurement in $\theta_{l}^{*}$ and $\phi_{l}^{*}$ in single top events \cite{ATLASangles}. 
A cut-based analysis is used to obtain a signal enriched data sample.
In order to account for efficiency and resolution effects, an analytic folding is performed to the simulated samples before comparing with the real data.
Two parameters are measured simultaneously: the fraction $f_{1}$ of decays containing transversely polarized W bosons and the phase $\delta_{-}$ between amplitudes for longitudinal and transversely polarized W bosons recoiling against left-handed b quarks, obtaining that $f_{1} = F_{\mathrm{R}} + F_{\mathrm{L}} = 0.37 \pm 0.05\,(\mathrm{stat.})\pm 0.05\,(\mathrm{syst.})$ and $\delta_{-} = -0.014\mathrm{\pi} \pm 0.023\mathrm{\pi}\,(\mathrm{stat.})\pm 0.028\mathrm{\pi}\,(\mathrm{syst.})$.
Limits to the real and imaginary parts of the ratio $g_{R}/V_{L}$ are extracted: $Re[g_{R}/V_{L}] \in [-0.36, 0.10]$ and $Im[g_{R}/V_{L}] \in [-0.17, 0.23]$ (Figure 4)

\begin{wrapfigure}{r}{0.49\textwidth}
 \vspace{-40pt}
   \begin{center}
     \includegraphics[width=0.49\textwidth]{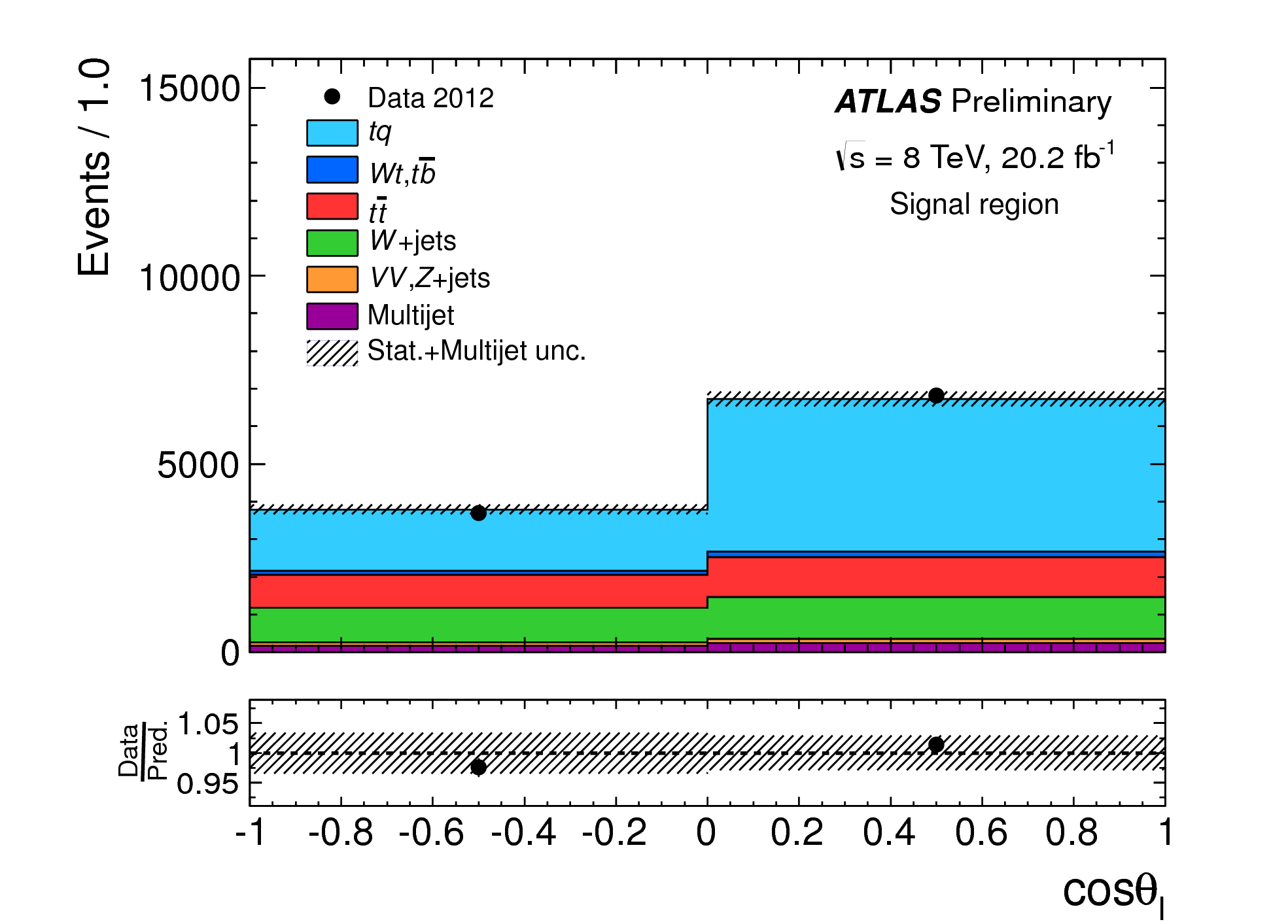}
    \end{center}
     \vspace{-20pt}
  \caption{$cos(\theta_{l})$ distribution in the signal region used to measure the $A_{FB}^{l}$ asymmetry. \cite{ATLAScou}}
 \vspace{-10pt}
\end{wrapfigure}

Another ATLAS study, using the 8 TeV dataset, has measured the top quark and W boson polarization observables from asymmetries in various angular distributions \cite{ATLAScou}.
The signal is enriched with cut-based techniques and later unfolded to parton level after subtracting the background contributions.
The angular asymmetries are extracted from the unfolded distributions. 
Of particular interest are the measurements of two asymmetries:
$A^{\rm{N}}_{\mathrm{FB}}$, which has the highest
sensitivity to the imaginary part of $g_{\mathrm{R}}$ \cite{AguilarBer}, and $A^{l}_{\mathrm{FB}}$, obtained from the $cos(\theta_{l})$ distribution (Figure 5), for which the CMS measurement shows a small tension with the SM prediction \cite{CMSpol}. 
This tension is not observed in the ATLAS analysis, the result of which is $A^{l}_{\mathrm{FB}} = 0.48 \pm 0.03\,(\mathrm{stat.})\pm 0.05\,(\mathrm{syst.})$, compatible with the SM. 
Limits to the imaginary part of $g_{\mathrm{R}}$ are obtained from these two asymmetries: $Im[g_{\mathrm{R}}] \in [-0.17, 0.06]$, assuming SM values for all other couplings.

Limits on anomalous Wtb couplings are obtained by CMS, using 7 and 8 TeV data \cite{CMScou}. 
The analysis is based on a multivariate discriminant.
A neural network (NN) has been used to diminish QCD background contamination. 
Multijet events are rejected with a dedicated Bayesian NN (BNN).
Only the muon channel has been used for the study.
After a preselection, 3 NN are defined for anomalous couplings, where two and three out of four anomalous couplings are considered simultaneously in two and three dimensional scenarios.
Limits are extracted from a simultaneous fit to the SM BNN and the anomalous BNNs outputs.
The obtained limits are the following: $V_{\mathrm{L}} > 0.98$,  $|V_{\mathrm{R}}| < 0.16$,  $|g_{\mathrm{L}}| < 0.57$  and  $ -0.049 < |g_{\mathrm{R}}| < 0.048$. 

 \vspace{-7.5pt}
\section{Top quark mass}

At hadron colliders, most top quark mass measurements are done using top pair samples, due to the higher cross-section of the strongly produced pairs, leading to larger available statistics.
Nevertheless, measuring the top quark mass with single top quark provides the following benefits:
\begin{itemize}
\setlength\itemsep{-0.25em}
\item The mass is obtained from a statistically independent sample.
\item There are less ambiguities in the jet-parton assignments, as there is only one b-jet in the final state, and thereby, less combinatorial background.
\item The production mechanisms are different, featuring a very different color flow. This is useful to check that there are no large systematic biases in t\=t measurements due to the modelling of non-perturbative QCD effects.
\item The typical energy scale of single top production is much smaller than the t\=t one.
\end{itemize}

ATLAS has measured the mass of the top quark in topologies enhanced with single top quarks produced in the t-channel using the 8 TeV dataset \cite{ATLASmass}.
The top quark processes are further enhanced using a neural network based discriminant. 
After the full event selection, non-top quark background fraction and t\=t events fraction are about 28\% and 26\% respectively.
A template method, using as estimator the distribution of the invariant mass of the lepton and b-tagged jet system, is used to extract the mass (Figure 6).
Using pseudo-experiments, a good linearity has been found between the input top quark mass and the mean value derived from the distribution of the reconstructed top quark mass. 
The final measurement is $m_{\mathrm{top}} = 172.2 \pm 0.7\,(\mathrm{stat.}) \pm 2.0\,(\mathrm{syst.})$ GeV.

Using the same channel, CMS has performed a cut-based analysis to measure the top quark mass with 8 TeV data \cite{CMSmass}.
Retaining only the positively charged muons and requiring the accompanying light jet to appear in forward regions, a signal to background ratio close to 3 is achieved.
The reconstructed top quark invariant mass distribution is fitted to extract the top quark mass. 
The fitting function is composed of multiple functions to take into account single top, t\=t pair and background events (Figure 7).
A calibration curve for the fit result is obtained from a set of simulated samples with different top quark mass. 
The final measurement is $m_{top} = 172.60 \pm 0.77\,(\mathrm{stat.}) ^{+ 0.97}_{- 0.93}\,(\mathrm{syst.})$ GeV.

\vspace{5pt}
\begin{centering}
\begin{minipage}[c]{0.477\textwidth}
\centering
\includegraphics[width=0.96\textwidth]{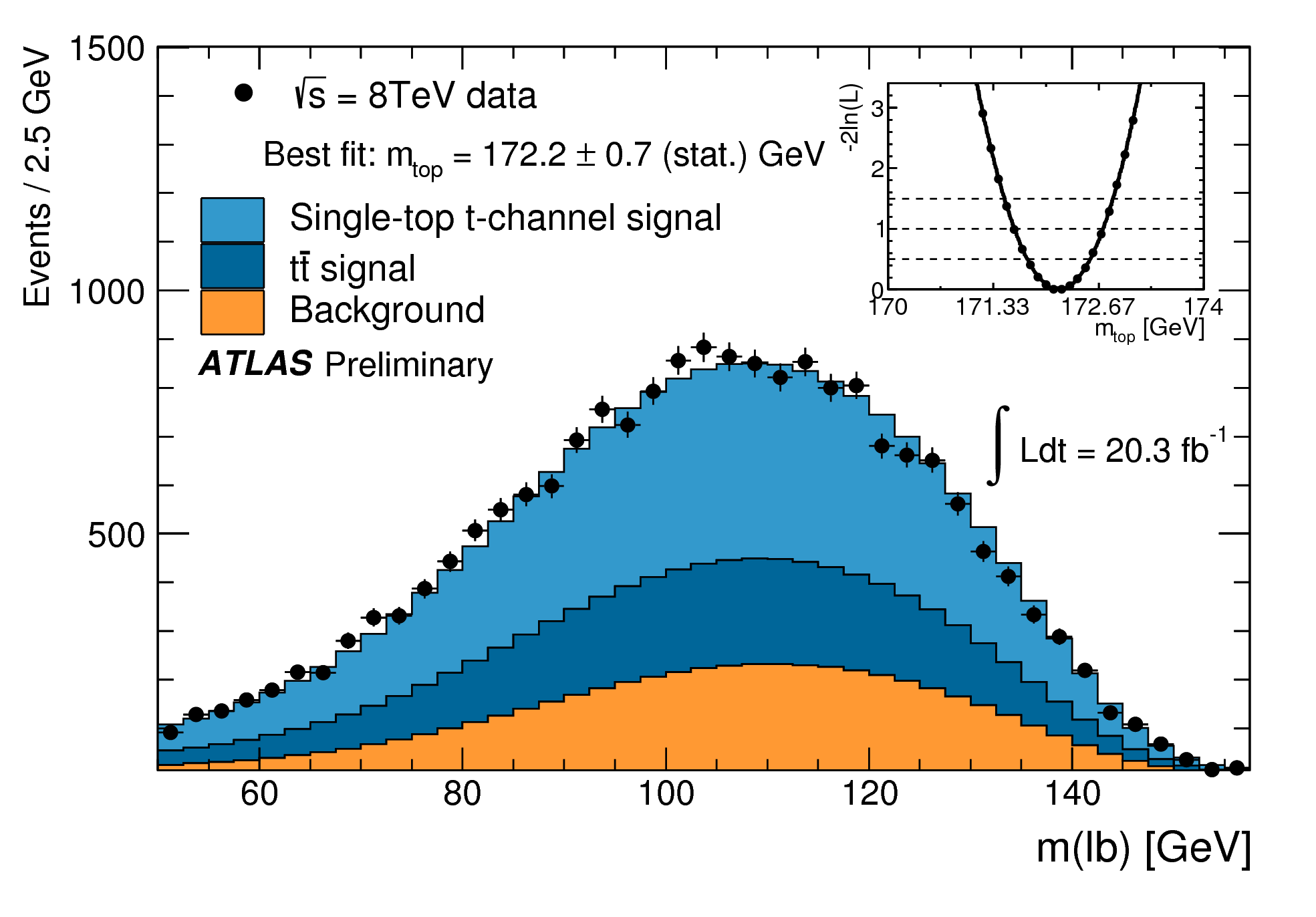}
\vspace{-5pt}
\captionof{figure}{Fitted m(lb) distribution in data with the normalization and $m_{top}$ being the best fit values. \cite{ATLASmass}}
\end{minipage}
\hspace{10pt}
\begin{minipage}[c]{0.47\textwidth}
\centering
\includegraphics[width=0.95\textwidth]{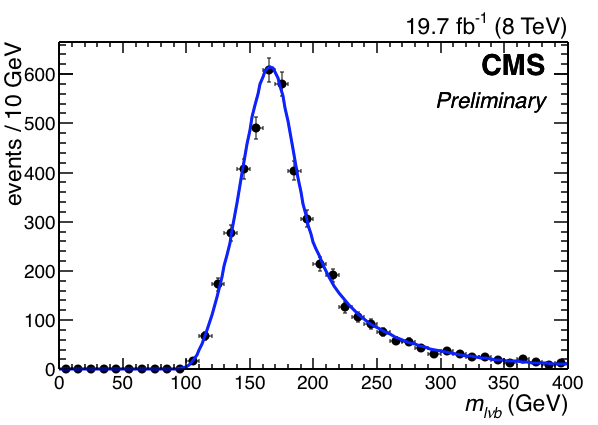}
\vspace{-5pt}
\captionof{figure}{Result of the fit in data to the top invariant mass shape. The solid blue line represents the fitted PDF.  \cite{CMSmass}}
\end{minipage}
\end{centering}

\section{Conclusions}

ATLAS and CMS collaborations have studied extensively the properties of the top quark in the single top quark production.
Searches for anomalous couplings in the Wtb vertex have been performed through the top quark and W boson polarizations. 
All measurements are in good agreement with the SM. 
There is a 2$\sigma$ effect in the top polarization measurement in t-channel, were the observed asymmetry is slightly less than the expected.
The top quark mass has been measured using single top events by ATLAS and CMS, achieving a comparable precision with measurements obtained with top pairs.

\end{document}